\documentclass[preprint,aps,eqsecnum,superscriptaddress,showpacs]{revtex4}
\usepackage{graphicx,amsfonts}
\begin{document}

\title{Noncommutative Correction to the Aharonov-Bohm Scattering: a
  Field Theory Approach}
\author{M. A. Anacleto}
\author{M. Gomes}
\author{ A. J. da Silva}
\affiliation{Instituto de F\'\i sica, Universidade de S\~ao Paulo\\
Caixa Postal 66318, 05315-970, S\~ao Paulo, SP, Brazil}
\email{anacleto, mgomes, ajsilva@fma.if.usp.br}
\author{D. Spehler}
\affiliation{Universit\'e Louis Pasteur, I. U. T.
All\'ee d'Ath\`enes, 67300 Schiltigheim, France}
\email{spehler@if.usp.br}
\date{\today}
\begin{abstract}
We study a noncommutative nonrelativistic theory in 2+1
dimensions of a scalar field coupled to the Chern-Simons field. 
In the commutative situation this model has been used to simulate the
Aharonov-Bohm effect in the field theory context.
We verified that,  contrarily to the commutative result, the
inclusion
of a quartic self-interaction of the scalar field is not necessary
to secure the ultraviolet renormalizability of the model. However, 
to obtain a smooth commutative limit the presence 
of a quartic gauge invariant self-interaction is required.
For small noncommutativity we fix the
corrections to the Aharonov-Bohm scattering and prove that up to
one-loop the model
is free from dangerous infrared/ultraviolet divergences.
\end{abstract}

\maketitle
\newpage
\pretolerance10000

\section{Introduction}
Noncommutative field theories present a series of unusual and
intriguing properties (see~\cite{Douglas} for some reviews). From a conceptual standpoint the inherent
nonlocality of these theories lead to an entanglement of scales so
that some ultraviolet (UV) divergences of their commutative counterparts
appear as infrared (IR) singularities. In general they are damaging to the
perturbative expansions although in some supersymmetric models
\cite{gomes2,bichl,gomes3} they could be under control. Noncommutative
field theories have also been used to clarify condensed matter
phenomena as the fractional Hall \cite{susskind} and Aharonov-Bohm (AB)
\cite{chaichian,Gamboa} effects.

In the context of nonrelativistic quantum mechanics, previous study on the 
noncommutative AB effect  have shown that, in contrast with the commutative
situation, the cross section for the scattering of scalar particles by a thin
solenoid does not vanish even when the magnetic field assumes certain
discrete values \cite{Gamboa}.

In this work we will proceed further the investigations on the changes
on the AB effect due to the noncommutativity of the space.  In our
study the effect will be simulated by a nonrelativistic field theory
describing spin zero particles interacting through a Chern-Simons (CS)
field. It is worth to recall that in the commutative scenario, to cancel
ultraviolet divergences and to obtain accordance with the exact
result, it was necessary to introduce a quartic self-interaction for
the scalar field \cite{lozano}. This result was reobtained by
considering the low momentum limit of the full relativistic theory \cite{boz}.
Even in the case of U(1) gauge symmetry to which we will restrict our
considerations, due to the noncommutativity, the CS
field is  similar to a  non-Abelian gauge field so that we will
be actually dealing with a non-Abelian AB effect \cite{lee}
 (see \cite{bergman}
for studies on the non-Abelian commutative AB effect using the CS field).
Besides, because of the change in the character of some divergences, from
ultraviolet to infrared, the renormalizability of the model may be in jeopardy.
However, in the present situation there are two possible orderings for
the quartic self-interactions. There is one more free
parameter and this could help to formulate a consistent model. In any case, as the limit of small noncommutativity is 
singular,  features different from those in~\cite{Gamboa} 
emerge from our analysis.

In this work all calculations are performed in the Coulomb gauge which
for nonrelativistic studies seems to be more adequate.  We show that
up to the one-loop order the UV divergences of the planar
contributions are canceled in the calculation of the four-point
function and, contrary to the commutative case, do not have a
conformal anomaly.  Hence, the planar part is renormalizable without
the contact interaction needed in the commutative situation.
Nevertheless, as mentioned before, the nonplanar part presents
logarithmic infrared divergences as the noncommutative parameter tends
to zero.  To eliminate these divergences we introduce in the
Lagrangian quartic interactions of the type $\frac{\lambda _{1}}{8}
[\phi ,\phi ^{\dagger }]_{\ast}\ast \lbrack \phi,\phi^{\dagger
}]_\ast$ and $\frac{\lambda _{2}}{8} \{\phi
,\phi^{\dagger}\}_\ast\ast\{\phi,\phi ^{\dagger }\}_\ast$, all field
products being Moyal ones. For general values of $\lambda_1$ and
$\lambda_2$ gauge invariance will be broken and UV divergences
originated from the quartic terms occur.  However, it turns out that,
for the special values $\lambda _{1}=\lambda _{2}=\lambda $, for which
the action is gauge invariant, these UV divergences are eliminated. We prove then
that IR divergences in the scattering amplitude disappear for 
special values of the coupling constant $\lambda$.

The paper is organized as follows. In section II, we introduce the
model, present its Coulomb gauge Feynman rules and
discuss some aspects of the renormalization program for the model. In section III, we compute the
particle-particle scattering up to order one-loop. 
We calculate the
scattering amplitude by separating the  planar and nonplanar parts and
complete  the one-loop analysis of the IR/UV divergences  initiated in
the previous section. Some integrals needed in the calculations are
collected in the Appendix.
Final comments are made in the Conclusions.
\section{Noncommutative Perturbative Theory}
We consider the noncommutative version of the theory of a 
nonrelativistic scalar field coupled with a CS
field in 2+1 dimensions described by the action 
\begin{eqnarray}
S[A,\phi ] &=&\int d^{3}x\left[ \frac{\kappa }
{2}\varepsilon ^{\mu \nu\lambda }
\left( A_{\mu }\ast \partial _{\nu }A_{\lambda }
+\frac{2ig}{3}A_{\mu }\ast A_{\nu }\ast A_{\lambda }\right)
-\frac{1}{2\xi }\partial_{i}A^{i}\ast \partial _{j}A^{j}\right.
\nonumber\\
&&+\left. i\phi ^{\dagger }\ast D_{t}\phi -\frac{1}{2m}
(\mathbf{D}\phi)^{\dagger }\ast (\mathbf{D}\phi )
-\frac{\lambda _{1}}{8}[\phi ,\phi^{\dagger }]_\ast
\ast \lbrack \phi ,\phi ^{\dagger }]_\ast\right.
\nonumber \\
&&-\left. \frac{\lambda _{2}}{8}\{\phi ,\phi ^{\dagger }\}_\ast
\ast \{\phi ,\phi^{\dagger }\}_\ast+\partial ^{i}\bar{c}\ast \partial _{i}c
+ig\partial ^{i}\bar{c}\ast[A_{i},c]\right],\label{accao} 
\end{eqnarray}

\noindent
where a Coulomb gauge fixing and the corresponding Faddeev-Popov terms are
already included.
The fields $\phi$ and $\phi ^{\dagger }$ belong to the
fundamental representation of the $U(1)$ gauge group
\begin{equation}
\phi\rightarrow (e^{i\Lambda})_{\ast}\ast\phi, 
\end{equation}
\begin{equation}
\phi^{\dagger}\rightarrow \phi^{\dagger}\ast(e^{-i\Lambda})_{\ast}, 
\end{equation}

\noindent
whereas the gauge field transforms as
\begin{equation}
A_{\mu }\rightarrow ({\textrm{e}}^{i\Lambda })_{\ast }\ast A_{\mu }\ast 
({\textrm{e}}^{-i\Lambda })_{\ast }+i[\partial _{\mu }
({\textrm{e}}^{i\Lambda })_{\ast }]\ast ({\textrm{e}}^{-i\Lambda })_{\ast }.
\label{4}
\end{equation}

\noindent
The covariant derivatives are given by 
\begin{eqnarray}
D_{t}\phi &=&\partial_{t}\phi+igA_{0}\ast\phi,  \nonumber\\
D_{i}\phi &=&\partial _{i}\phi+igA_{i}\ast\phi.
\end{eqnarray}

\noindent
Notice that there are two different orderings  for the quartic
self-interaction. In (\ref{accao}) they were written 
in terms of Moyal  commutators and anticommutators of the scalar
fields. 

\begin{figure}
\centering
\scalebox{0.7}{\includegraphics{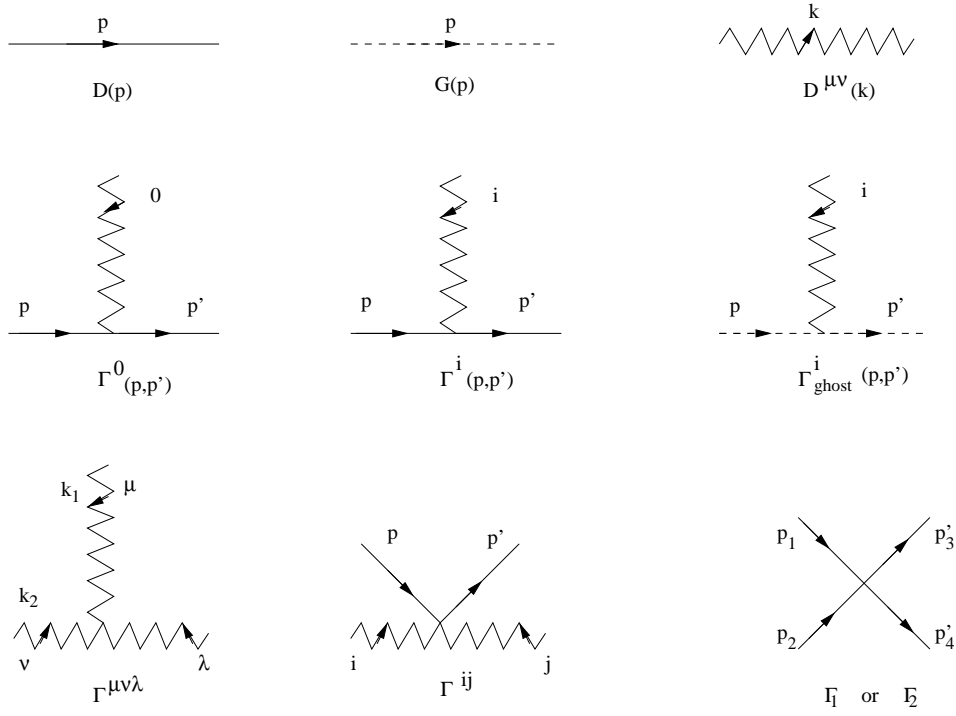}}
\caption{{ Feynman rules for the action (\ref{accao}). }}
\label{regrab}
\end{figure}

For convenience, we will work in a strict Coulomb gauge obtained
by letting $\xi \rightarrow 0$.
Furthermore, we will use a graphical notation where the CS field, the
matter field and the ghost field propagators are represented by
wavy, continuous and dashed lines respectively.
The graphical representation for the Feynman rules is given
in Fig. \ref{regrab} and the corresponding analytical expression are:

(i) The matter field propagator:
\begin{equation}
D(p)=\frac{i}{p_{0}-\frac{\mathbf{p}^2}{2m} + i\epsilon},
\end{equation}

(ii) The ghost field propagator: 
\begin{equation}
G(p)=-\frac{i}{\mathbf{p}^2},
\end{equation}

(iii) The gauge field propagator in the limit $\xi \rightarrow 0$ is 
\begin{equation}
D_{\mu \nu }(k)=\frac{\varepsilon _{\mu \nu \lambda }\bar{k}^{\lambda }}{%
\kappa \mathbf{k}^{2}},
\end{equation}

where $\bar{k}^{\lambda}$=$(0,{\bf k})$.

(iv) The analytical expressions associated with the vertices are:
\begin{eqnarray}
&&\Gamma ^{0}(p,p^{\prime}) =-ige^{ip\theta p^{\prime}}, \\
&&\Gamma ^{i}(p,p^{\prime})=\frac{ig}{2m}(p+p^{\prime})^{i}
e^{ip\theta p^{\prime}}, \\
&&\Gamma ^{i}_{ghost}(p,p^{\prime})=-2gp^{\prime i}
\sin(p\theta p^{\prime}), \\
&&\Gamma ^{\mu\nu\lambda}(k_{1},k_{2}) =2ig\kappa \varepsilon ^{\mu \nu
\lambda }\sin (k_{1}\theta k_{2}), \\
&&\Gamma ^{ij}(k_{1},k_{2},p,p^{\prime}) =-\frac{ig^{2}}{m}
\cos(k_{1}\theta k_{2})e^{ip\theta p^{\prime}}\delta ^{ij}, \\
&&\Gamma _{1}(p_{1},p_{3}^{\prime},p_{2},p_{4}^{\prime})=i\lambda _{1}
[\sin (p_{1}\theta p_{3}^{\prime})\sin(p_{2}\theta p_{4}^{\prime})
+\sin(p_{1}\theta p_{4}^{\prime})\sin (p_{2}\theta p_{3}^{\prime})], 
\\
&&\Gamma _{2}(p_{1},p_{3}^{\prime},p_{2},p_{4}^{\prime})=-i\lambda _{2}
[\cos(p_{1}\theta p_{3}^{\prime})\cos(p_{2}\theta p_{4}^{\prime}) 
+\cos(p_{1}\theta p_{4}^{\prime})\cos(p_{2}\theta p_{3}^{\prime})]. 
\end{eqnarray}
In these expressions we have defined
$k_{1}\theta k_{2}\equiv \frac{1}{2}\theta ^{\mu\nu}k_{1\mu}k_{2\nu}$,
where $\theta^{\mu\nu}$ is the anti-symmetric matrix which characterize
the noncommutativity of the underlying space. For simplicity we assume
that $\theta^{0i}=0$ and $\theta^{ij}=\theta \epsilon^{ij}$ with
$\epsilon^{ij}$ being the two dimensional Levi-Civit\`a symbol,
normalized as $\epsilon^{12}=1$.

In the one-loop approximation there are  quadratic divergences,
associated with the two point functions of the gauge and scalar
fields, linear divergences, associated with the scalar field four
point function and logarithmic divergences, associated with the
three point functions $<A_\mu \phi \phi^\dag>$. In the sequel we shall
analyze each one of these divergences.  

\begin{figure}
\centering
\scalebox{0.7}{\includegraphics{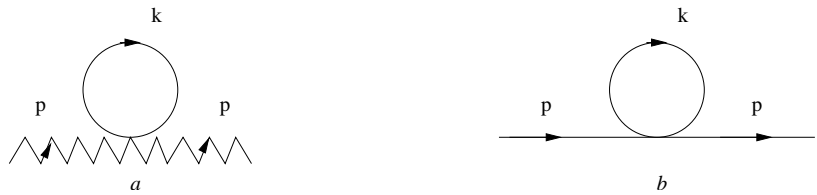}}
\caption{{ One-loop contributions to the gauge and scalar field propagators.}}
\label{scalar}
\end{figure}

(1) Gauge and  scalar fields two-point functions. The graph in
Fig. \ref{scalar}$a$
which contributes to the   gauge field two point function is 
planar so that it can be eliminated by an adequate counterterm.
Specifically, the only one-loop nonvanishing contribution is
given by
\begin{eqnarray}
\Pi_{a}^{ij}&=&-\frac{ig^{2}\delta^{ij}}{2m}\int \frac{d^{2}{\bf{k}}}
{(2\pi)^{2}}=-\frac{ig^{2}\delta^{ij}\Lambda^{2}}{8\pi m}.
\end{eqnarray}

\noindent 
This is a gauge noninvariant term and  shall be removed by a 
$A_i A_i$ counterterm so that gauge (BRST) 
invariance remains unbroken. 

The diagram in Fig \ref{scalar}$b$ which contributes to the scalar 
field two-point
function have both   planar and  nonplanar parts. As before, the
planar part can be eliminated by a counterterm. For general
values of $\lambda_1$ and $\lambda_2$ , the nonplanar
part although ultraviolet finite may generate nonintegrable infrared
singularities. These nonplanar parts are however canceled if one
chooses $\lambda_1=\lambda_2$ which is also the condition to enforce gauge 
invariance.

\begin{figure}
\centering
\scalebox{0.7}{\includegraphics{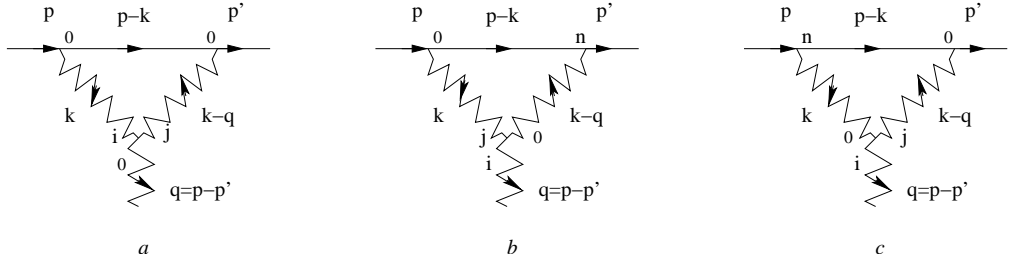}}
\caption{{ One-loop contributions to the three point vertex function. 
The numerals correspond to the indices of the gauge field propagator.}}
\label{fig3}
\end{figure}

(2) As Lorentz invariance is broken, the  three point
    function $<TA_\mu \phi\phi^\dag>$  presents two types of
    divergences:

\noindent
(a) The one-loop contribution to  $<TA_0 \phi\phi^\dag>$, drawn in 
Fig. \ref{fig3}$a$  
is given by
\begin{eqnarray}
\Gamma^{0}&=-&\frac{g^{3}e^{ip\theta p^{\prime}}}{2\kappa}
\lim_{\mu\rightarrow 0}\int\frac{d^{2}{\bf{k}}}{(2\pi)^{2}}
\frac{({\bf{q}}\wedge{\bf{k}})[1-e^{-2iq\theta k}]}
{{\bf{k}}^{2}[({\bf{k}}-{\bf{q}})^{2}+\mu^{2}]},\label{f1}
\end{eqnarray}

\noindent
where we introduced the parameter $\mu$ to regulate possible infrared
divergences in the intermediary steps of the calculation. For small
$\theta$ we obtain 
\begin{eqnarray}
\Gamma^{0}&=&\frac{ig^{3}\theta{\bf{q}}^{2}e^{ip\theta p^{\prime}}}{8\pi\kappa}
\left[\ln\left(\frac{\theta{\bf{q}}^{2}}{2}\right)+\gamma -1\right],
\end{eqnarray}

\noindent
where $\gamma$ is the 
Euler-Mascheroni constant. Notice that $\Gamma^0$ is  finite in the infrared limit. 

\noindent
(b) Concerning the three point function $<TA^i \phi\phi^\dag>$, 
we
found two contributions
\begin{eqnarray}
&&\Gamma^{i}_{1}=\frac{g^{3}e^{ip\theta p^{\prime}}}{2m\kappa}
\lim_{\mu\rightarrow 0}\int\frac{d^{2}{\bf{k}}}{(2\pi)^{2}}
\frac{[({\bf{p}}-{\bf{k}})\wedge {\bf{p}}^{\prime}]
[1-e^{-2iq\theta k}]}{{\bf{k}}^{2}[({\bf{k}}-{\bf{q}})^{2}+\mu^{2}]}k^{i},\\
&&\Gamma^{i}_{2}=-\frac{ g^{3}e^{ip\theta p^{\prime}}}{2m\kappa}
\lim_{\mu\rightarrow 0}\int\frac{d^{2}{\bf{k}}}{(2\pi)^{2}}
\frac{({\bf{p}}\wedge{\bf{k}})
[1-e^{-2iq\theta k}]}{{\bf{k}}^{2}[({\bf{k}}-{\bf{q}})^{2}+\mu^{2}]}(k^{i}-q^{i}),
\end{eqnarray}

\noindent
associated with the graphs in the Figs. \ref{fig3}$b$  and \ref{fig3}$c$,
respectively. For small $\theta$ the calculation of these amplitudes furnishes
the following results for their planar and nonplanar parts

(b1) Planar parts:
\begin{eqnarray}
&&\Gamma^{i}_{planar\,1}=-\frac{g^{3}e^{ip\theta p^{\prime}}}{8\pi m\kappa}
\left\{q^{i}\frac{{\bf{p}}\wedge {\bf{p}}^{\prime}}{{\bf{q}}^{2}}
\ln\left(\frac{\mu^{2}}{{\bf{q}}^{2}}\right)
+\varepsilon^{in}p^{\prime{n}}
\left[\ln\left(\frac{\mu^{2}}{\Lambda^2}\right)-2\right]\right\},\\
&&\Gamma^{i}_{planar\,2}=-\frac{g^{3}e^{ip\theta p^{\prime}}}{8\pi m\kappa}
\varepsilon^{in}p^{n}
\left[\ln\left(\frac{\Lambda^{2}}{{\bf{q}}^{2}}\right)
+2\right].
\end{eqnarray}

(b2) Nonplanar parts:
\begin{eqnarray}
\!\!\!\!\!\Gamma^{i}_{nplanar\,1}&=&\frac{g^{3}e^{ip\theta p^{\prime}}}{8\pi m\kappa}
\left\{({\bf{p}}\wedge {\bf{p}}^{\prime})\left[\frac{q^{i}}{{\bf{q}}^{2}}
\ln\left(\frac{\mu^{2}}{{\bf{q}}^{2}}\right)
-i{\tilde{q}}^{i}
\left(\ln\left(\frac{\theta{\bf{q}}^{2}}{2}\right)
+\gamma-1\right)\right]\right.
\nonumber\\
&&+\left. \varepsilon^{in}p^{\prime{n}}
\left[\ln\left(\frac{\mu^{2}}{{\bf{q}}^{2}}\right)
+\ln\left(\frac{\theta{\bf{q}}^{2}}{2}\right)+1+\gamma\right]\right\},
\\
\!\!\!\!\!\Gamma^{i}_{nplanar\,2}&=&\!\!-\frac{g^{3}e^{ip\theta p^{\prime}}}
{8\pi m\kappa}
\left\{[\varepsilon^{in}p^{n}\!+i\theta q^{i}({\bf{p}}.{\bf{q}})]
\left[\ln\left(\frac{\theta{\bf{q}}^{2}}{2}\right)
\!+\gamma\right]\!
+\! [\varepsilon^{in}p^{n}-i\theta q^{i}({\bf{p}}.{\bf{q}})]\right\}\!,
\end{eqnarray}
where we have defined ${\tilde{q}}^{i}\equiv \theta^{ij}q_{j}$.

Summing up these parts we get the total contribution for small
$\theta$
\begin{eqnarray}
\Gamma^{i}&=&\Gamma^{i}_{planar}+\Gamma^{i}_{nplanar}
\nonumber\\
&=&-\frac{ig^{3}e^{ip\theta p^{\prime}}}{8\pi m\kappa}
[{\tilde{q}}^{i}({\bf{p}}\wedge {\bf{p}}^{\prime})
+\theta q^{i}({\bf{p}}.{\bf{q}})]
\left[\ln\left(\frac{\theta{\bf{q}}^{2}}{2}\right)+\gamma -1\right]
\nonumber\\
&&-\frac{g^{3}e^{ip\theta p^{\prime}}\varepsilon^{in}q^{n}}
{8\pi m\kappa}\left[\ln\left(\frac{\Lambda^{2}\theta}{2}\right)
+\gamma+3\right].\label{f2}
\end{eqnarray}

Notice that the final results do not depend on $\mu$. The infrared
divergences being only logarithmic are harmless whereas the
ultraviolet
divergence has to be eliminated by a counterterm. It remains to
analyze
the four point function but that will be done in the next section together
with the computation of the two body scattering matrix.

\section{Particle-Particle Scattering}

\begin{figure}
\centering
\scalebox{0.7}{\includegraphics{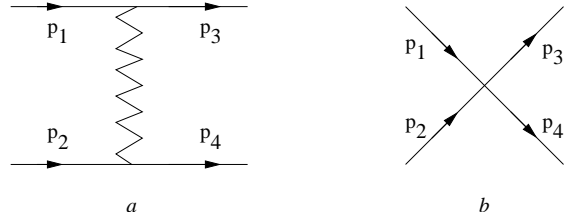}}
\caption{{ Tree level scattering.}}
\label{treelevel}
\end{figure}

\begin{figure}
\centering
\scalebox{0.7}{\includegraphics{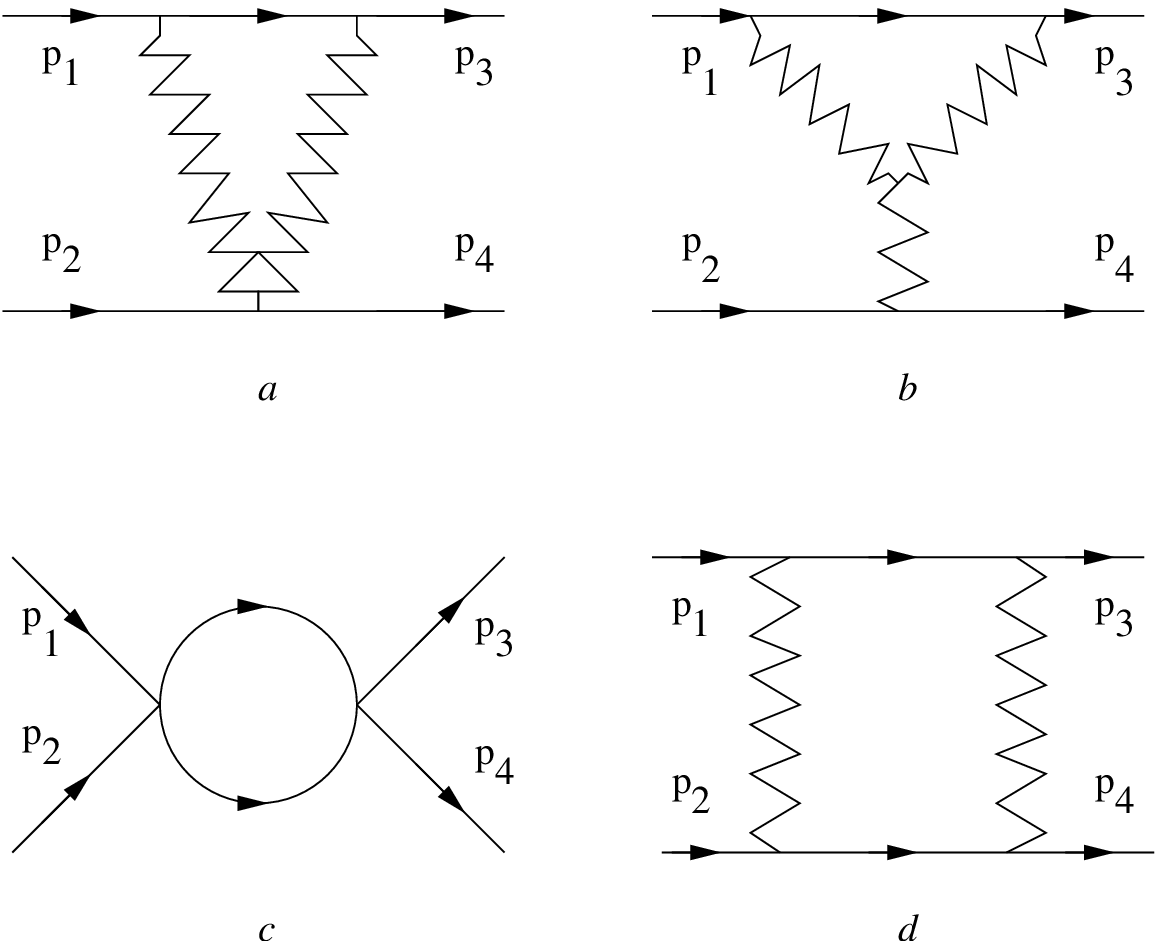}}
\caption{ One-loop scattering.} 
\label{umloop}
\end{figure}

The object that  we wish to analyze is the four point function associated with
the scattering of two identical particles in the
center-of-mass frame. The relevant diagrams are depicted in
Figs. \ref{treelevel} and \ref{umloop} but for sake of simplicity we
have drawn only the $s$-channel processes.
In the tree approximation the gauge
part of the two  body
scattering amplitude  is given by (see Fig. \ref{treelevel}$a$), 
\begin{equation}
{\cal{A}}_{a}^{0}(\varphi)=-\frac{2ig^2({\bf{p}}_{1}\wedge {\bf{p}}_{3})}
{m\kappa}\left[\frac{e^{i(p_{1}\theta p_{3}+p_{2}\theta p_{4})}}
{({\bf{p}}_{1}-{\bf{p}}_{3})^{2}}
-\frac{e^{-i(p_{1}\theta p_{3}+p_{2}\theta p_{4})}}
{({\bf{p}}_{1}+{\bf{p}}_{3})^{2}}\right],\label{espa}
\end{equation}
where ${\bf{p}}_{1}$, ${\bf{p}}_{2}$  and ${\bf{p}}_{3}$, ${\bf{p}}_{4}$ are the incoming and outgoing momenta.
Since  $\theta_{ij}=\theta\varepsilon_{ij}$, the phase is
\begin{eqnarray}
p_{1}\theta p_{3}+p_{2}\theta p_{4}=\theta({\bf{p}}_{1}\wedge {\bf{p}}_{3})
=\theta{\bf{p}}^{2}\sin\varphi=\bar{\theta}\sin\varphi,
\end{eqnarray} 
where we have defined  $\bar{\theta}\equiv \theta{\bf{p}}^{2}$,
${\bf{p}}^{2}
\equiv {\bf{p}}^{2}_{1}={\bf{p}}^{2}_{3}$ and
$\varphi $ is the scattering angle. 
Therefore, Eq. (\ref{espa}) can be rewritten as 
\begin{equation}
{\cal{A}}_{a}^{0}(\varphi)=-\frac{ig^2}{m\kappa}
\left[\frac{e^{i\bar{\theta}\sin\varphi}}
{1-\cos\varphi}
-\frac{e^{-i\bar{\theta}\sin\varphi}}
{1+\cos\varphi}\right]\sin\varphi,
\end{equation}

\noindent
which for small $\bar \theta$ gives ${\cal{A}}_{a}^{0}(\varphi)\approx
 -\frac{2ig^2}{m\kappa}(\cot\varphi+ i\bar \theta$).

By taking into account the quartic self-interaction we have
the additional contribution
\begin{eqnarray}
{\cal{A}}_{b}^{0}(\varphi)&=&\lambda _{1}
[\sin(p_{1}\theta p_{3})\sin (p_{2}\theta p_{4})
+\sin(p_{1}\theta p_{4})\sin (p_{2}\theta p_{3})] 
\nonumber\\
&&-\lambda _{2}[\cos (p_{1}\theta p_{3})\cos(p_{2}\theta p_{4})
+\cos(p_{1}\theta p_{4})\cos(p_{2}\theta p_{3})]
\nonumber\\
&=&2\lambda _{1}\sin^{2}\left(\frac{\bar{\theta}\sin\varphi}{2}\right)
-2\lambda_{2}\cos^{2}\left(\frac{\bar{\theta}\sin\varphi}{2}\right),
\end{eqnarray}

\noindent
coming from the graph in Fig. \ref{treelevel}$b$.
Thus, the full tree level amplitude   is 
\begin{eqnarray}
{\cal{A}}(\varphi)&=&-\frac{ig^{2}}{m\kappa}
\left[\cot\left(\frac{\varphi}{2}\right)e^{i\bar{\theta}\sin\varphi}
-\tan\left(\frac{\varphi}{2}\right)e^{-i\bar{\theta}\sin\varphi}\right]
\nonumber \\
&&+2\lambda_{1}\sin^{2}\left(\frac{\bar{\theta}\sin\varphi}{2}\right)
-2\lambda_{2}\cos^{2}\left(\frac{\bar{\theta}\sin\varphi}{2}\right).
\end{eqnarray}

The one-loop contribution to the scattering amplitude is depicted in the
Fig. \ref{umloop} (all other possible one-loop graphs vanish). 
The analytic expressions associated with these graphs, after
performing the $k_{0}$ integration, are

1. For the triangle graph shown in Fig. \ref{umloop}$a$:
\begin{eqnarray}
{\cal{A}}_{a}(\varphi)&=&-\frac{g^{4}}{4m\kappa ^{2}}
e^{i(p_{1}\theta p_{3}+p_{2}\theta p_{4})}
\int\frac{d^{2}{\bf{k}}}{(2\pi)^{2}}\frac{{\bf{k}}.({\bf{k}}-{\bf{q}})}
{{\bf{k}}^{2}({\bf{k}}-{\bf{q}})^{2}}\left[1+e^{-2iq\theta k}\right] 
\nonumber\\
&&+(p_{1} \leftrightarrow p_{2})+(p_{3}\leftrightarrow p_{4})+ 
\left(
p_{1}\leftrightarrow p_{2} \quad \mbox{and}\quad 
p_{3}\leftrightarrow p_{4}
\right),
\end{eqnarray}
where ${\bf{q}}={\bf{p}}_{1}-{\bf{p}}_{3}$ is the momentum transferred,

2. For the trigluon graph shown in Fig. \ref{umloop}$b$
 (${\bf{q}}^{\prime}={\bf{p}}_1+{\bf{p}}_3$):
\begin{equation}
{\cal{A}}_{b}(\varphi)={\cal{A}}^{1}_{b}(\varphi)
+{\cal{A}}^{2}_{b}(\varphi)+{\cal{A}}^{3}_{b}(\varphi),
\end{equation}
where
\begin{eqnarray}
{\cal{A}}^{1}_{b}(\varphi)&=&\frac{g^{4}}{4m\kappa ^{2}}
e^{i(p_{1}\theta p_{3}+p_{2}\theta p_{4})}\int \frac{d^{2}{\bf{k}}}{(2\pi)^{2}}
\left[\frac{{\bf{k}}^{2}{\bf{q}}^{2}-({\bf{k}}.{\bf{q}})^{2}
+({\bf{k}}.{\bf{q}}^{\prime})({\bf{k}}.{\bf{q}})
-({\bf{k}}.{\bf{q}}^{\prime}){\bf{q}}^{2}}
{{\bf{k}}^{2}{\bf{q}}^{2}({\bf{k}}-{\bf{q}})^{2}}\right]
\nonumber\\
&&\times\left[1-e^{-2iq\theta k}\right]
+(p_{1} \leftrightarrow p_{2})+(p_{3}\leftrightarrow p_{4})+ 
(p_{1}\leftrightarrow p_{2} 
\quad \mbox{and}\quad 
p_{3}\leftrightarrow p_{4}),
\nonumber\\
{\cal{A}}^{2}_{b}(\varphi)&=&\frac{g^{4}}{4m\kappa ^{2}}
e^{i(p_{1}\theta p_{3}+p_{2}\theta p_{4})}
\int \frac{d^{2}{\bf{k}}}{(2\pi)^{2}}
\left[\frac{{\bf{k}}^{2}{\bf{q}}^{2}
-2({\bf{k}}.{\bf{q}})({\bf{k}}.{\bf{p}}_{1})}
{{\bf{k}}^{2}{\bf{q}}^{2}({\bf{k}}-{\bf{q}})^{2}}\right]
\left[1-e^{-2iq\theta k}\right]
\nonumber\\
&&+(p_{1} \leftrightarrow p_{2})+(p_{3}\leftrightarrow p_{4})+ 
(p_{1}\leftrightarrow p_{2} 
\quad \mbox{and}\quad 
p_{3}\leftrightarrow p_{4}),
\nonumber\\
{\cal{A}}^{3}_{b}(\varphi)&=&-\frac{g^{4}}{4m\kappa ^{2}}
e^{i(p_{1}\theta p_{3}+p_{2}\theta p_{4})}
\int \frac{d^{2}{\bf{k}}}{(2\pi)^{2}}
\left[\frac{({\bf{k}}.{\bf{q}}^{\prime}){\bf{q}}^{2}}
{{\bf{k}}^{2}{\bf{q}}^{2}({\bf{k}}-{\bf{q}})^{2}}\right]
\left[1-e^{-2iq\theta k}\right]
\nonumber\\
&&+(p_{1} \leftrightarrow p_{2})+(p_{3}\leftrightarrow p_{4})+ 
(p_{1}\leftrightarrow p_{2} 
\quad \mbox{and}\quad 
p_{3}\leftrightarrow p_{4}),
\end{eqnarray}

3. For the bubble graph shown in Fig. \ref{umloop}$c$:
\begin{eqnarray}
{\cal{A}}_{c}(\varphi)&=&\int\frac{ d^{2}{\bf{k}}}{(2\pi)^{2}}
\left\{4m(\lambda _{1}-\lambda_{2})^{2}
-8m(\lambda _{1}^{2}-\lambda_{2}^{2})
\cos(2k\theta p_{1})\right.
\nonumber\\
&&+\left.2m(\lambda _{1}+\lambda_{2})^{2}[\cos(2k\theta q) 
+\cos(2k\theta q^{\prime})]\right\}
\frac{1}{({\bf{k}}^{2}-{\bf{p}}^{2}-i\epsilon)}.
\end{eqnarray}

The above integrals being logarithmically divergent need a regularization.
Thus, although not indicated, a cutoff regularization is being implicitly
assumed.

4. For the box graph in Fig. \ref{umloop}$d$: 
\begin{equation}
{\cal{A}}_{d}(\varphi)={\cal{A}}_{d}^{1}(\varphi)
+{\cal{A}}_{d}^{2}(\varphi),
\end{equation}
\noindent
where
\begin{equation}
{\cal{A}}_{d}^{1}(\varphi )=\frac{4g^{4}}{m\kappa ^{2}}
\int \frac{d^{2}{\bf{k}}}{(2\pi)^{2}}\frac{({\bf{p}}_{1}\wedge
{\bf{k}})({\bf{p}}_{3}\wedge 
{\bf{k}})
e^{2iq\theta k}}{({\bf{k}}-{\bf{p}}_{1})^{2}
({\bf{k}}-{\bf{p}}_{3})^{2}({\bf{k}}^{2}-{\bf{p}}^{2}-i\epsilon)},
\label{1}
\end{equation}
\begin{equation}
{\cal{A}}_{d}^{2}(\varphi)=-\frac{4g^{4}}{m\kappa ^{2}}
\int\frac{d^{2}{\bf{k}}}{(2\pi)^{2}}\frac{({\bf{p}}_{1}\wedge 
{\bf{k}})({\bf{p}}_{3}\wedge {\bf{k}})
e^{-2iq^{\prime}\theta k}}{({\bf{k}}+{\bf{p}}_{1})^{2}
({\bf{k}}-{\bf{p}}_{3})^{2}({\bf{k}}^{2}-{\bf{p}}^{2}-i\epsilon)}.
\label{2}
\end{equation}

To  compute the above integrals, we separate their
planar and nonplanar contributions. A simplifying
aspect is that the box graph  is purely nonplanar.

\subsection{Planar Contribution}

In the perturbative expansion there is one planar contribution containing
phase factors which depend only on the external momenta.
Although the interaction induced by the noncommutativity is nonlocal, the
divergences in the momentum integration for closed internal loops are the
same as for the commutative theory. 

The calculations of the planar contributions are standard
so that we just list the results:

1. The  planar part of the triangle graph,
\begin{eqnarray}
\label{tria}
{\cal{A}}_{a}^{p}(\varphi)&=&-\frac{g^{4}}{4m\kappa ^{2}}
\int \frac{d^{2}{\bf{k}}}{(2\pi)^{2}}
\left[\frac{{\bf{k}}.({\bf{k}}-{\bf{q}})e^{i\bar{\theta}\sin\varphi}}
{{\bf{k}}^{2}({\bf{k}}-{\bf{q}})^{2}}
+\frac{{\bf{k}}.({\bf{k}}+{\bf{q}}^{\prime})e^{-i\bar{\theta}\sin\varphi}}
{{\bf{k}}^{2}({\bf{k}}+{\bf{q}}^{\prime})^{2}}\right.
\nonumber\\
&&+\left. ({\bf{q}}\rightarrow -{\bf{q}},
\quad{\bf{q}}^{\prime}\rightarrow -{\bf{q}}^{\prime})\right],
\end{eqnarray}

\noindent
gives
\begin{eqnarray}
{\cal{A}}_{a}^{p}(\varphi)&=&-\frac{g^{4}}{4\pi m\kappa ^{2}}
[\cos(\bar{\theta}\sin\varphi)\ln \left(\frac{\Lambda ^{2}}
{{\bf{p}}^{2}}\right) \nonumber\\
&&-\ln |2\sin (\varphi /2)|e^{i\bar{\theta}\sin\varphi}
-\ln|2\cos (\varphi /2)|e^{-i\bar{\theta}\sin\varphi}].
\end{eqnarray}

2. The planar part of the trigluon graph is more intricate being
   given by
\begin{equation}
{\cal{A}}_{b}^{p}(\varphi)={\cal{A}}_{b}^{1p}(\varphi)
+{\cal{A}}_{b}^{2p}(\varphi)+{\cal{A}}_{b}^{3p}(\varphi),
\end{equation}
where
\begin{eqnarray}
{\cal{A}}_{b}^{1p}(\varphi)&=&\frac{g^{4}}{4m\kappa ^{2}}
\int \frac{d^{2}{\bf{k}}}{(2\pi)^{2}}
\left[\frac{({\bf{k}}^{2}{\bf{q}}^{2}-({\bf{k}}.{\bf{q}})^{2})
e^{i\bar{\theta}\sin\varphi}}
{{\bf{k}}^{2}{\bf{q}}^{2}({\bf{k}}-{\bf{q}})^{2}}
+\frac{({\bf{k}}^{2}{\bf{q}}^{\prime2}-({\bf{k}}.{\bf{q}}^{\prime})^{2})
e^{-i\bar{\theta}\sin\varphi}}
{{\bf{k}}^{2}{\bf{q}}^{\prime2}({\bf{k}}-{\bf{q}}^{\prime})^{2}}\right]
\nonumber\\
&&+ ({\bf{q}}\rightarrow -{\bf{q}},
\quad{\bf{q}}^{\prime}\rightarrow -{\bf{q}}^{\prime}),
\end{eqnarray}
\begin{eqnarray}
{\cal{A}}_{b}^{2p}(\varphi)&=&\frac{g^{4}}{4m\kappa ^{2}}
\int \frac{d^{2}{\bf{k}}}{(2\pi)^{2}}
\left[\frac{{\bf{k}}^{2}{\bf{q}}^{2}-2({\bf{k}}.{\bf{q}})
({\bf{p}}_{1}.{\bf{k}})}
{{\bf{k}}^{2}{\bf{q}}^{2}({\bf{k}}-{\bf{q}})^{2}}
e^{i\bar{\theta}\sin\varphi}
+\frac{{\bf{k}}^{2}{\bf{q}}^{\prime2}-2({\bf{k}}.{\bf{q}}^{\prime})
({\bf{p}}_{2}.{\bf{k}})}
{{\bf{k}}^{2}{\bf{q}}^{\prime2}({\bf{k}}-{\bf{q}}^{\prime})^{2}}
e^{-i\bar{\theta}\sin\varphi}\right]
\nonumber\\
&&+ ({\bf{q}}\rightarrow -{\bf{q}},
\quad{\bf{q}}^{\prime}\rightarrow -{\bf{q}}^{\prime}),
\end{eqnarray}
\begin{eqnarray}
{\cal{A}}_{b}^{3p}(\varphi)&=&-\frac{g^{4}}{4m\kappa ^{2}}
\int \frac{d^{2}{\bf{k}}}{(2\pi)^{2}}
\left[\left(\frac{({\bf{k}}.{\bf{q}}^{\prime})}
{{\bf{k}}^{2}({\bf{k}}-{\bf{q}})^{2}}
-\frac{({\bf{k}}.{\bf{q}}^{\prime})}
{{\bf{k}}^{2}({\bf{k}}+{\bf{q}})^{2}}\right)
e^{i\bar{\theta}\sin\varphi}\right.
\nonumber\\
&&+\left. \left(\frac{({\bf{k}}.{\bf{q}})}
{{\bf{k}}^{2}({\bf{k}}-{\bf{q}}^{\prime})^{2}}
-\frac{({\bf{k}}.{\bf{q}})}
{{\bf{k}}^{2}({\bf{k}}+{\bf{q}}^{\prime})^{2}}\right)
e^{-i\bar{\theta}\sin\varphi}\right].
\end{eqnarray}

\noindent
and the final result is
\begin{eqnarray}
{\cal{A}}_{b}^{p}(\varphi)&=&\frac{g^{4}}{4\pi m\kappa ^{2}}
[\cos(\bar{\theta}\sin\varphi)
\left[\ln\left(\frac{\Lambda ^{2}}{{\bf{p}}^{2}}\right) +1\right]
\nonumber \\
&&-\ln |2\sin (\varphi /2)|e^{i\bar{\theta}\sin\varphi}
-\ln|2\cos (\varphi /2)|e^{-i\bar{\theta}\sin\varphi}].
\end{eqnarray}

Notice now that the sum of the planar contribution,
${\cal{A}}_{a}^{p}(\varphi)$ and ${\cal{A}}_{b}^{p}(\varphi)$, is
\begin{equation}
{\cal{A}}_{a+b}^{p}(\varphi)=\frac{g^{4}}{4\pi m\kappa^{2}}
\cos(\bar{\theta}\sin\varphi),
\end{equation}

\noindent
so that the divergent parts of these graphs mutually cancel,
unlike in the commutative case~\cite{bergman}. Thus, to eliminate
the ultraviolet divergences a quartic self-interaction does
not seem to be necessary. However, we should be cautious because as remarked
before some ultraviolet divergences have been transmuted into
infrared ones so that the quartic self-interaction may still
be needed.

3. The contribution of planar part of the bubble graph is
   logarithmically divergent and  is equal to
\begin{eqnarray}
{\cal{A}}_{c}^{p}(\varphi)=\frac{4m(\lambda _{1}-\lambda_{2})^{2}}
{(2\pi)^{2}}
\int d^{2}{\bf{k}}\frac{1}{{(\bf{k}}^{2}-{\bf{p}}^{2}-i\epsilon)}
=\frac{m(\lambda _{1}-\lambda_{2})^{2}}{\pi}
\left[\ln\left(\frac{\Lambda^{2}}{{\bf{p}}^{2}}\right)+i\pi\right].
\label{cont}
\end{eqnarray}

We can get rid of the  divergence by setting
$\lambda _{1}=\lambda _{2}=\lambda $. The  total planar part of the
amplitude  is therefore
\begin{eqnarray}
\label{partial}
{\cal{A}}_{\mbox{1-loop}}^{p}(\varphi)&=&-\frac{ig^{2}}{m\kappa}
[\cot (\varphi /2)e^{i\bar{\theta}\sin\varphi}
-\tan(\varphi/2)e^{-i\bar{\theta}\sin\varphi}] 
\nonumber\\
&&-2\lambda\cos(\bar{\theta}\sin\varphi)
+\frac{g^{4}}{4\pi m\kappa ^{2}}\cos(\bar{\theta}\sin\varphi),
\end{eqnarray}

\noindent
furnishing up to first order in the parameter
$\bar{\theta}$,
\begin{eqnarray}
{\cal{A}}_{\mbox{1-loop}}^{p}(\varphi)
&=&-\frac{ig^{2}}{m\kappa}\left[\cot\left(\frac{\varphi}{2}\right)
-\tan \left(\frac{\varphi}{2}\right)+i\bar{\theta}\sin\varphi
\left[\cot\left(\frac{\varphi}{2}\right) 
+\tan \left(\frac{\varphi}{2}\right) \right] \right] 
\nonumber\\
&&-2\lambda +\frac{g^{4}}{4\pi m\kappa^{2}}, 
\nonumber\\
&=&-\frac{i2g^{2}}{m\kappa}
\left[\cot\varphi+i\bar\theta\right] 
-2\lambda +\frac{g^{4}}{4\pi m\kappa ^{2}}.
\end{eqnarray}

\subsection{Nonplanar Contribution}

The nonplanar contributions are given by terms which contain extra
phase factors depending on the internal (loop) momenta.
For the graphs (\ref{umloop}$a$) these contributions are
\begin{eqnarray}
\label{trianp}
{\cal{A}}_{a}^{np}(\varphi)&=&-\frac{g^{4}}{4m\kappa ^{2}}
e^{i\bar{\theta}\sin\varphi}\int\frac{d^{2}{\bf{k}}}{(2\pi)^{2}}
\frac{{\bf{k}}.({\bf{k}}-{\bf{q}})}{{\bf{k}}^{2}({\bf{k}}-{\bf{q}})^{2}}
e^{-2iq\theta k} 
\nonumber\\
&&+(p_{1} \leftrightarrow p_{2})+(p_{3}\leftrightarrow p_{4})+ 
(p_{1}\leftrightarrow p_{2} 
\quad \mbox{and}\quad 
p_{3}\leftrightarrow p_{4}),
\end{eqnarray}

Let us begin by computing the first term in the r.h.s. of the above
expression. This is done straightforwardly by using Feynman
parameterization and the result \cite{Gel}
\begin{equation}
\int\frac{d^{n}k}{(2\pi)^{n}}\frac{e^{ik_{\alpha}\tilde{p}^{\alpha }}}
{[k^{2}-M^{2}]^{\lambda}}=i(-1)^{\lambda}
\frac{M^{n/2-\lambda}}{2^{\lambda-1}(2\pi)^{n/2}\Gamma[\lambda]}
\frac{K_{n/2-\lambda}\left(\sqrt{-M^{2}{\tilde{p}}^{2}}\right)}
{(-{\tilde{p}^{2}})^{n/2-\lambda}},
\label{I-basica-pura}
\end{equation}

\noindent 
where $K_{\nu}$ is the modified Bessel function of order
$\nu$. Proceeding in this way we obtain
\begin{equation}
{\cal{A}}_{a1}^{np}(\varphi)=-\frac{g^4}{4\pi m\kappa^2}
\int_{0}^{1}dx\left[K_{0}(\sqrt{a^2\bar{\theta}^2})
-\sqrt{a^2\bar{\theta}^2}K_{1}(\sqrt{a^2\bar{\theta}^2})\right]
e^{i\bar{\theta}\sin\varphi},
\end{equation}

\noindent
where $a^2=16 x(1-x)\sin^{4}(\varphi/2)$. Collecting this with the
results for the other terms then provides
\begin{eqnarray}
{\cal{A}}_{a}^{np}(\varphi)&=&-\frac{g^4}{4\pi m\kappa^2}
\int_{0}^{1}dx\left\{\left[K_{0}(\sqrt{a^2\bar{\theta}^2})
-\sqrt{a^2\bar{\theta}^2}K_{1}(\sqrt{a^2\bar{\theta}^2})\right]
e^{i\bar{\theta}\sin\varphi}\right.
\nonumber\\
&&+\left.\left[K_{0}(\sqrt{b^2\bar{\theta}^2})
-\sqrt{b^2\bar{\theta}^2}K_{1}(\sqrt{b^2\bar{\theta}^2})\right]
e^{-i\bar{\theta}\sin\varphi}\right\},
\end{eqnarray}
 
\noindent 
with $b^2=16x(1-x)\cos^{4}(\varphi/2)$.

Let us turn now to the computation of the nonplanar part of the  graph
with the trigluon vertex. We have
\begin{equation}
{\cal{A}}_{b}^{np}(\varphi)={\cal{A}}_{b}^{1np}(\varphi)
+{\cal{A}}_{b}^{2np}(\varphi)+{\cal{A}}_{b}^{3np}(\varphi),
\end{equation}
where
\begin{eqnarray}
{\cal{A}}_{b}^{1np}(\varphi)&=&-\frac{g^{4}}{4m\kappa ^{2}}
e^{i\bar{\theta}\sin\varphi}\int \frac{d^{2}{\bf{k}}}{(2\pi)^{2}}
\left[\frac{{\bf{k}}^{2}{\bf{q}}^{2}-({\bf{k}}.{\bf{q}})^{2}
+({\bf{k}}.{\bf{q}}^{\prime})({\bf{k}}.{\bf{q}})
-({\bf{k}}.{\bf{q}}^{\prime}){\bf{q}}^{2}}
{{\bf{k}}^{2}{\bf{q}}^{2}({\bf{k}}-{\bf{q}})^{2}}\right]
e^{-2iq\theta k}
\nonumber\\
&&+(p_{1} \leftrightarrow p_{2})+(p_{3}\leftrightarrow p_{4})+ 
(p_{1}\leftrightarrow p_{2} 
\quad \mbox{and}\quad 
p_{3}\leftrightarrow p_{4}),
\nonumber\\
{\cal{A}}_{b}^{2np}(\varphi)&=&-\frac{g^{4}}{4m\kappa ^{2}}
e^{i\bar{\theta}\sin\varphi}\int \frac{d^{2}{\bf{k}}}{(2\pi)^{2}}
\left[\frac{{\bf{k}}^{2}{\bf{q}}^{2}
-2({\bf{k}}.{\bf{q}})({\bf{k}}.{\bf{p}}_{1})}
{{\bf{k}}^{2}{\bf{q}}^{2}({\bf{k}}-{\bf{q}})^{2}}\right]
e^{-2iq\theta k}
\nonumber\\
&&+(p_{1} \leftrightarrow p_{2})+(p_{3}\leftrightarrow p_{4})+ 
(p_{1}\leftrightarrow p_{2} 
\quad \mbox{and}\quad 
p_{3}\leftrightarrow p_{4}),
\nonumber\\
{\cal{A}}_{b}^{3np}(\varphi)&=&\frac{g^{4}}{4m\kappa ^{2}}
e^{i\bar{\theta}\sin\varphi}\int \frac{d^{2}{\bf{k}}}{(2\pi)^{2}}
\left[\frac{({\bf{k}}.{\bf{q}}^{\prime}){\bf{q}}^{2}}
{{\bf{k}}^{2}{\bf{q}}^{2}({\bf{k}}-{\bf{q}})^{2}}\right]
e^{-2iq\theta k}
\nonumber\\
&&+(p_{1} \leftrightarrow p_{2})+(p_{3}\leftrightarrow p_{4})+ 
(p_{1}\leftrightarrow p_{2} 
\quad \mbox{and}\quad 
p_{3}\leftrightarrow p_{4}).
\end{eqnarray}

We calculate these contributions by following the same steps
described for the previous case.
Thus, we obtain 
\begin{eqnarray}
{\cal{A}}_{b}^{np}(\varphi)&=&-\frac{g^4}{4\pi m\kappa^2}
\int_{0}^{1}dx\left[[(3+2i\bar{\theta}\sin\varphi)
K_{0}(\sqrt{a^2\bar{\theta}^2})\right.
\nonumber\\
&&-\left.\sqrt{a^2\bar{\theta}^2}K_{1}(\sqrt{a^2\bar{\theta}^2})]
e^{i\bar{\theta}\sin\varphi}+[(3-2i\bar{\theta}\sin\varphi)
K_{0}(\sqrt{b^2\bar{\theta}^2})\right.
\nonumber\\
&&-\left.\sqrt{b^2\bar{\theta}^2}K_{1}(\sqrt{b^2\bar{\theta}^2})]
e^{-i\bar{\theta}\sin\varphi}\right],
\end{eqnarray}

\noindent
which for small $\bar \theta$ behaves as
\begin{eqnarray}
{\cal{A}}_{a+b}^{np}(\varphi)&=&\left[\ln\left(\frac{\bar{\theta}}
{2}\right)+\gamma\right]\left[\frac{2g^{4}}{\pi m\kappa ^{2}}\right]
+\frac{2g^{4}}{\pi m\kappa ^{2}}\ln (2\sin\varphi) 
\nonumber\\
&&+i\frac{2\bar{\theta}\sin\varphi g^{4}}{\pi m\kappa ^{2}}
\ln[\tan(\varphi/2)]
+\frac{2g^{4}}{\pi m\kappa ^{2}} 
+O({\bar{\theta}}^{2}).
\end{eqnarray}

The nonplanar contribution of the bubble graph is 
\begin{eqnarray}
{\cal{A}}_{c}^{np}(\varphi)&=&\frac{m}{(2\pi)^{2}}
\int d^{2}{\bf{k}}
\left[\frac{2(\lambda _{1}+\lambda_{2})^{2}[\cos(2k\theta q) 
+\cos(2k\theta q^{\prime})]}{{(\bf{k}}^{2}-{\bf{p}}^{2})}\right.
\nonumber\\           
&&-\left.\frac{8(\lambda _{1}^{2}-\lambda_{2}^{2})
\cos(2k\theta p_{1})}{{(\bf{k}}^{2}-{\bf{p}}^{2})}\right].
\end{eqnarray}
 By integrating over the internal momenta we get
\begin{equation}
{\cal{A}}_{c}^{np}(\varphi)=\frac{m}{\pi}
[(\lambda _{1}+\lambda _{2})^{2}[K_{0}(i2\sin(\varphi/2)\bar{\theta})
+K_{0}(i2\cos(\varphi/2)\bar{\theta})]
-4(\lambda _{1}^{2}-\lambda _{2}^{2})K_{0}(i\bar{\theta})],
\end{equation}
which, for small  $\bar{\theta}$,  is given by
\begin{eqnarray}
{\cal{A}}_{c}^{np}(\varphi)&=&-
\left[\ln\left(\frac{\bar{\theta}}{2}\right) 
+\gamma \right]\left[\frac{2m}{\pi} (\lambda _{1}+\lambda_{2})^{2}
-\frac{4m}{\pi}(\lambda _{1}^{2}-\lambda_{2}^{2})\right]
-\frac{m}{\pi }(\lambda _{1}+\lambda _{2})^{2}\ln[2\sin\varphi]
\nonumber\\
&&-i m (\lambda _{1}+\lambda_{2})^{2}
+2i m(\lambda _{1}^{2}-\lambda_{2}^{2}).
\end{eqnarray}

\noindent
Setting $\lambda _{1}=\lambda _{2}=\lambda $, that,  as remarked
before,  eliminates the ultraviolet divergence of the planar part
of the same graph, yields
\begin{equation}
{\cal{A}}_{c}^{np}(\varphi)=-\frac{8m\lambda ^{2}}{\pi}
\left[\ln\left(\frac{\bar{\theta}}{2}\right)+\gamma\right]
-\frac{4m\lambda ^{2}}{\pi}\ln[2\sin\varphi]-4i m\lambda^{2}.
\end{equation}

For small $\theta$ the amplitudes (\ref{1}) and (\ref{2}) associated
with the box graph are
\begin{equation}
{\cal{A}}_{d}^{1}(\varphi)=\frac{4g^{4}}{m\kappa ^{2}}
\int\frac{d^{2}{\bf{k}}}{(2\pi)^{2}}
\frac{({\bf{p}}_{1}\wedge {\bf{k}})({\bf{p}}_{3}\wedge {\bf{k}})
[1+i\theta ({\bf{p}}_{1}
-{\bf{p}}_{3})\wedge {\bf{k}]}}
{({\bf{k}}-{\bf{p}}_{1})^{2}({\bf{k}}-{\bf{p}}_{3})^{2}
({\bf{k}}^{2}-{\bf{p}}^{2}-i\epsilon)},
\end{equation}
and
\begin{equation}
{\cal{A}}_{d}^{2}(\varphi)=-\frac{4g^{4}}{m\kappa ^{2}}
\int \frac{d^{2}{\bf{k}}}{(2\pi)^{2}}\frac{({\bf{p}}_{1}\wedge 
{\bf{k}})({\bf{p}}_{3}\wedge {\bf{k}})
[1-i\theta ({\bf{p}}_{1}+{\bf{p}}_{3})\wedge 
{\bf{k}}]}{({\bf{k}}+{\bf{p}}_{1})^{2}({\bf{k}}-{\bf{p}}_{3})^{2}
({\bf{k}}^{2}-{\bf{p}}^{2}-i\epsilon)}.
\end{equation}

The $\theta$ independent part of these expressions give 
\begin{eqnarray}
{\cal{A}}^{1}_{d}(\varphi)\vert_{\theta=0}
&=&-\frac{g^{4}}{m\kappa ^{2}}\int_{0}^{\Lambda^2}\frac{d{\bf{k}}^{2}}
{(2\pi )^{2}}\frac{[I_{2}-\cos (\varphi )I_{0}]}{({\bf{k}}^{2}
-{\bf{p}}^{2}-i\epsilon )}\nonumber \\
&=&-\frac{g^{4}}{4\pi m\kappa ^{2}}
\left[2\ln\left(2\sin \left( \varphi /2\right) \right) +i\pi \right]
,\\
{\cal{A}}^{2}_{d}(\varphi)\vert_{\theta=0}
&=&-\frac{g^{4}}{m\kappa ^{2}}\int_{0}^{\Lambda^2}\frac{d{\bf{k}}^{2}}
{(2\pi )^{2}}\frac{[{\cal I}_{2}-\cos (\varphi ){\cal I}_{0}]}{({\bf{k}}^{2}
-{\bf{p}}^{2}-i\epsilon )}\nonumber \\
&=&-\frac{g^{4}}{4\pi m\kappa ^{2}}
\left[2\ln\left(2\cos \left( \varphi /2\right) \right) +i\pi \right],
\end{eqnarray}

\noindent
where $I_{0}$, $I_{2}$, ${\cal I}_{0}$ and  ${\cal I}_{2}$ are
defined in the appendix A.  Summing the above results we get
\begin{equation}
{\cal A}_d(\varphi)\vert_{\theta=0}={\cal{A}}^{1}_{d}(\varphi)
\vert_{\theta=0}+{\cal{A}}^{2}_{d}(\varphi)\vert_{\theta=0}
=-\frac{g^{4}}{2\pi m\kappa ^{2}}\left[\ln(2\sin\varphi)+i\pi\right].
\end{equation}

Concerning the terms proportional to $\theta$ we have
\begin{eqnarray}
{\cal{A}}_{\theta d}^{1}(\varphi)&=&\frac{4i\theta g^{4}}{m\kappa ^{2}}
\int\frac{d^{2}{\bf{k}}}{(2\pi)^{2}}
\frac{({\bf{p}}_{1}\wedge {\bf{k}})({\bf{p}}_{3}\wedge {\bf{k}})
({\bf{p}}_{1}-{\bf{p}}_{3})\wedge {\bf{k}}}
{({\bf{k}}-{\bf{p}}_{1})^{2}({\bf{k}}-{\bf{p}}_{3})^{2}
({\bf{k}}^{2}-{\bf{p}}^{2}-i\epsilon)} 
\nonumber\\
&=&-\frac{i\theta g^{4}\sin(\varphi/2)}{m\kappa ^{2}}
\int_{0}^{\Lambda^2}\frac{d{\bf{k}}^{2}}{(2\pi)^{2}}
\frac{|{\bf{k}}||{\bf{p}}|[I_{1}-2\cos(\varphi)I_{1}+I_{3}]}
{({\bf{k}}^{2}-{\bf{p}}^{2}-i\epsilon)}\nonumber\\
&=&-\frac{i\bar{\theta}g^{4}\sin\varphi}{2\pi m\kappa^{2}}
[i\pi+1+2\ln\left[ 2\sin(\varphi/2)\right], 
\end{eqnarray}

\noindent
and
\begin{eqnarray}
{\cal{A}}_{\theta d}^{2}(\varphi)&=&\frac{4i\theta g^{4}}{m\kappa ^{2}}
\int\frac{d^{2}{\bf{k}}}{(2\pi)^{2}}
\frac{({\bf{p}}_{1}\wedge 
{\bf{k}})({\bf{p}}_{3}\wedge {\bf{k}})({\bf{p}}_{1}+{\bf{p}}_{3})
\wedge {\bf{k}}}{({\bf{k+p}}_{1})^{2}({\bf{k}}-{\bf{p}}_{3})^{2}
({\bf{k}}^{2}-{\bf{p}}^{2}-i\epsilon)}, 
\nonumber\\
&=&\frac{i\theta g^{4}\cos(\varphi/2)}{m\kappa ^{2}}
\int_{0}^{\Lambda^2}\frac{d{\bf{k}}^{2}}{(2\pi)^{2}}
\frac{|{\bf{k}}||{\bf p}|[I_{3}^{\prime}-I_{1}^{\prime}
-2\cos(\varphi)I_{1}^{\prime}]}{({\bf{k}}^{2}
-{\bf{p}}^{2}-i\epsilon)},
\nonumber\\
&=&\frac{i\bar{\theta}g^{4}\sin\varphi}
{2\pi m\kappa ^{2}}[i\pi+1+2\ln\left[2\cos(\varphi /2)\right] ].
\end{eqnarray}

Adding these contributions, we obtain
\begin{eqnarray}
{\cal{A}}_{\theta d}(\varphi)&=&{\cal{A}}_{\theta d}^{1}(\varphi)
+{\cal{A}}_{\theta d}^{2}(\varphi)
\nonumber \\
&=&-\frac{i\bar{\theta}g^{4}\sin\varphi}{\pi m\kappa ^{2}}
\ln\left[\tan\left(\frac{\varphi}{2}\right)\right], 
\end{eqnarray}

\noindent
Therefore, the total amplitude for the box graph  is finite and, up to order 
$\bar{\theta}$, is given by
\begin{equation}
{\cal{A}}_{d}(\varphi)=-\frac{g^{4}}{2\pi m\kappa ^{2}}
\left[\ln(2\sin\varphi) +i\pi \right]
-\frac{i\bar{\theta}g^{4}\sin\varphi}{\pi m\kappa ^{2}}
\ln\left[\tan\left(\frac{\varphi}{2}\right)\right]. 
\end{equation}

Summing all the contributions, we get the total one-loop amplitude
\begin{eqnarray}
{\cal{A}}_{\mbox{1-loop}}(\varphi)
&=&{\cal{A}}_{\mbox{1-loop}}^{p}(\varphi)
+{\cal{A}}_{\mbox{1-loop}}^{np}(\varphi)+{\cal{A}}_{d}(\varphi ) 
\nonumber\\
&=&-\frac{2ig^{2}}{m\kappa}\cot\varphi
+\frac{2\bar{\theta}g^{2}}{m\kappa}-2\lambda 
-\frac{ig^{4}}{2m\kappa ^{2}}-4i m\lambda^{2}
+\frac{9g^{4}}{4\pi m\kappa ^{2}} 
\nonumber\\
&&+\left(\frac{2g^4}{\pi m\kappa^2}-\frac{8m\lambda^2}{\pi}\right)
\left[\ln\left(\frac{\bar\theta}{2}\right)+\gamma\right] 
+\left(\frac{3g^{4}}{2\pi m\kappa ^{2}}
-\frac{4m\lambda ^{2}}{\pi}\right)\ln [2\sin\varphi]
\nonumber\\
&& +\frac{i\bar{\theta}g^{4}\sin\varphi}{\pi m\kappa ^{2}}
\ln\left[\tan\left(\frac{\varphi}{2}\right)\right] + O(\bar{\theta}^{2}).
\end{eqnarray}
                 
Notice the logarithmic singularity at $\bar \theta=0$. This is an
example of the
aforementioned transmutation of  ultraviolet singularities into
infrared ones. Had we used $\theta$ just as a regularization parameter
then {\it a fortiori} we should remove such singularity which implies
that $\lambda= \pm \frac{g^2}{2m\kappa}$.

\section{Conclusions}

In this work we studied the nonrelativistic and noncommutative theory
of scalar particles minimally coupled to a CS field and also
subject to a quartic self-interaction. In opposition to the
commutative case, the ultraviolet renormalizability of the model does
not require the presence of the quartic self-interaction of the
scalar field. However, the inclusion of a gauge invariant
self-interaction is obligatory if a smooth commutative limit is
demanded. In fact, the complete elimination of both ultraviolet and
infrared singularities only occurs for a critical value, 
$\lambda =$$\pm \frac{g^{2}}{2m\kappa}$, of the gauge invariant 
quartic self-interaction.
For small values of $\theta$ there are
corrections which modify qualitative and quantitative
aspects of the commutative  AB effect, as it should be expected due to
the nonlocal character of the noncommutative interaction.
In the tree approximation and to first order in the noncommutative 
parameter the correction to the two body  scattering is
isotropic. This is in qualitative accord with the results of
holonomy calculations  ~\cite{chaichian,Gamboa}. 
However, in various aspects our result differs
from \cite{Gamboa}. For example, except for the special values of the quartic
self-coupling, $\lambda =$$\pm \frac{g^{2}}{2m\kappa}$, our scattering amplitude
is not analytical for small $\theta$. Furthermore, for small scattering
angle $\varphi$, the noncommutative correction found by us shows a 
$\varphi \ln\varphi$ dependence.  These features are not present in
\cite{Gamboa} and may be traced to the use of different formalisms. 
In fact, due to the inherent nonlocality of the noncommutative 
situation different results may arise from the use of otherwise equivalent
procedures \cite{Gamboa1}.

\setcounter{equation}{0}
\renewcommand{\theequation}{\Alph{section}.\arabic{equation}}
\appendix
\section{Integrals}

We evaluate the following integrals
\begin{eqnarray}
I_{n} &=&\int_{0}^{2\pi}d\alpha \frac{\cos (n\alpha )}
{[2\cos(\alpha-\varphi /2)-f][2\cos (\alpha +\varphi /2)-f]} 
\nonumber\\
&=&\frac{2\pi }{B_{+}\sin (\varphi /2)\sqrt{f^{2}-4}}\left[ Z_{-}^{n-1}
\sin\left[ (n+1)\frac{\varphi}{2}\right] 
+Z_{-}^{n+1}\sin \left[ (n-1)\frac{\varphi }{2}\right] \right],
\end{eqnarray}
\begin{eqnarray}
{\cal{I}}_{n}&=&\int_{0}^{2\pi}d\alpha\frac{\cos(n\alpha)}
{[2\cos(\alpha -\varphi /2)-f][2\cos(\alpha +\varphi /2)+f]} 
\nonumber\\
&=&-\frac{\pi \left[ 1+(-1)^{n}\right]}{2\cos(\varphi/2)B_{-}
\sqrt{f^{2}-4}}\left[Z_{-}^{n-1}\cos\left[(n+1)\frac{\varphi}{2}\right]
+Z_{-}^{n+1}\cos\left[(n-1)\frac{\varphi }{2}\right] \right],
\end{eqnarray} 
\begin{eqnarray}
I_{n}^{\prime}&=&\int_{0}^{2\pi}d\alpha \frac{\sin (n\alpha)}
{[2\cos(\alpha -\varphi /2)-f][2\cos (\alpha +\varphi /2)+f]}
\label{sn}
\nonumber\\
&=&-\frac{\pi \left[1-(-1)^{n}\right]}
{\cos (\varphi/2)B_{-}\sqrt{f^{2}-4}}\left[ Z_{-}^{n-1}
\sin \left[(n+1)\frac{\varphi }{2}\right] 
+Z_{-}^{n+1}\sin\left[(n-1)\frac{\varphi}{2}\right]\right],
\end{eqnarray}

where 
\begin{eqnarray}
&&Z=\exp (i\alpha ) \quad {\mbox{and}} \quad  W=\exp (i\varphi /2),
\\
&&Z_{\pm}=\frac{1}{2}\left[ f\pm \sqrt{f^{2}-4}\right],
\\
&&B_{\pm}=f^{2}-2(1 \pm \cos\varphi),
\quad
f=\frac{{\bf{k}}^{2}+{\bf{p}}^{2}}{|{\bf{k}}||{\bf{p}}|}.
\end{eqnarray}

For $n=0,1,2$ and $3$ these formula furnish
\begin{eqnarray}
\label{Io}
&&I_{0}=\frac{2\pi f}{B_{+}\sqrt{f^{2}-4}}, 
\quad {\cal{I}}_{0}=-\frac{2\pi f}{B_{-}\sqrt{f^{2}-4}},
\quad I_{0}^{\prime}=0, 
\\
&&I_{1}=\frac{2\cos (\varphi /2)I_{0}}{f}, 
\quad {\cal{I}}_{1}=0, \quad 
I_{1}^{\prime }=-\frac{4\pi\sin\left(\varphi/2\right)}
{B_{-}\sqrt{f^{2}-4}}, 
\\
&&I_{2}=I_{0}-\frac{\pi f}{\sqrt{f^{2}-4}}+\pi, \quad 
{\cal{I}}_{2}=-{\cal{I}}_{0}-\frac{\pi f}{\sqrt{f^{2}-4}}+\pi, \quad
I_{2}^{\prime}=0, 
\label{I2}\\
&&I_{3}=\left[1+B_{+}-\frac{B_{+}f^{2}}{2}\right]I_{1}
+2\pi\cos(\varphi/2)f, \quad 
{\cal{I}}_{3}=0, 
\nonumber\\
&&I_{3}^{\prime}=-\left[1+B_{-}-\frac{B_{-}f^{2}}{2}\right]I_{1}^{\prime}
+2\pi\sin(\varphi/2)f.
\nonumber\\
\end{eqnarray}

\newpage



\begin{thebibliography}{99}
\bibitem{Gamboa1} Even if one stays at the quantum mechanical level, the
special definition for the Moyal product in polar coordinates adopted
in \cite{Gamboa} may originate contributions not present with
the usual definition.
\end{thebibliography}
\end{document}